\documentclass[aps, prd,  twocolumn, preprintnumbers,nofootinbib,10pt,eqsecnum]{revtex4-1}
\pdfoutput=1
\usepackage{myoptions}

\usepackage{bbm}
\usepackage[caption=false]{subfig}

\usepackage{chngcntr}
\counterwithout{equation}{section}

\begin{document}

\title{Membrane theory of entanglement dynamics from holography}

\author{M\'ark Mezei}
\affiliation{
Simons Center for Geometry and Physics, SUNY, Stony Brook, NY 11794, USA}

\begin{abstract}

\noindent Recently, a minimal membrane description of the entanglement dynamics of large regions in generic chaotic systems was conjectured in~\cite{Jonay:2018yei}. Analytic results in random circuits and spin chain numerics support this theory. We show that the results found by the author in~\cite{Mezei:2016zxg} about the dynamics of entanglement entropy in theories with a holographic dual can be reformulated in terms of the same minimal membrane, providing strong evidence that the membrane theory describes all chaotic systems. We discuss the implications of our results for tensor network approaches to holography and the holographic renormalization group.

\end{abstract}

\maketitle`

\section{Introduction}

The dynamics of entanglement entropy in closed quantum systems out of equilibrium is among the most fundamental aspects of thermalization, which has been a subject of recent intense activity in many branches of physics~\cite{Calabrese:2005in,bardarson2012unbounded,serbyn2013universal,Kim:2013bc,kaufman2016quantum}. One may wonder, if there exists an effective theory describing this process. Chaotic systems unitarily evolving from a homogenous short-range entangled high energy initial state are expected to ``erase the memory" of the fine details of the initial state at the local thermalization timescale $t_\text{loc}$, hence universality may emerge in the scaling limit 
\es{ScalingLimit}{
R,t\gg t_\text{loc}\,,
}
where $R$ is the characteristic length scale of the subregion $A$ whose entanglement entropy $S[A(t)]$ we want to determine. We imagine that similar to the gradient expansion of hydrodynamics, the effective theory will have the expansion parameter ${t_\text{loc}/ R}$. In this paper, we will study strongly interacting theories, where $t_\text{loc}\sim \beta$, where $\beta$ is the inverse temperature. Henceforth we will replace $t_\text{loc}$ with $\beta$.  The quasiparticle model of entanglement growth~\cite{Calabrese:2005in} is often used in the literature as an effective model, but it has recently become clear that, apart from special setups and integrable models, it cannot  reproduce the entanglement dynamics of chaotic systems~\cite{Balasubramanian:2011at,Asplund:2013zba,Asplund:2015eha,Casini:2015zua}.

Holographic duality is a powerful framework to study real time dynamics of certain strongly coupled chaotic quantum field theories in $d$ spacetime dimensions. On the gravitational side of the duality, the computation of entanglement entropy reduces to extremization of a codimension-2 surface in $d+1$-dimensional Lorentzian geometry~\cite{Ryu:2006bv,Ryu:2006ef,Hubeny:2007xt}, and entropy growth was studied in many important papers~\cite{AbajoArrastia:2010yt,Albash:2010mv,Balasubramanian:2010ce,Balasubramanian:2011ur,Balasubramanian:2011at,Hartman:2013qma,Liu:2013iza,Liu:2013qca}. In~\cite{Mezei:2016zxg}, the author simplified the holographic prescription in the scaling limit \eqref{ScalingLimit} even further, which enabled the analytic computation of the entropy in some symmetric setups. 

Recently, based on earlier analytic results  in random circuits~\cite{Nahum:2016muy} and spin chain numerics, Jonay, Huse, and Nahum 
proposed a minimal membrane theory of entropy growth in chaotic systems~\cite{Jonay:2018yei}. In this model, the entropy $S[A(t)]$  is given by the energy of a minimal codimension-1 membrane with angle dependent tension ${\cal E}\le(v\ri)$, where $v$ stands for the angle with the vertical direction,  stretching between two faces of a slab of $d$-dimensional Minkowski spacetime of height $t$.  The lower face of the slab represents the initial state and the membrane end perpendicularly to it, while the upper face represents the timeslice at $t$ and the boundary condition for the membrane is that it is anchored to the entangling surface $\p A(t)$.

In this paper, we show that the holographic results of~\cite{Mezei:2016zxg} can be exactly reformulated in this language, by partially solving the holographic extremization problem which reduces it to a membrane minimization problem. This result provides strong evidence for the membrane theory of entropy growth, and unifies high energy and condensed matter approaches to entropy dynamics. Holography enables us to prove various properties of the membrane tension ${\cal E}\le(v\ri)$, and allows for a determination of ${\cal E}\le(v\ri)$ in terms of final equilibrium state properties encoded in the dual equilibrium black brane geometry. ${\cal E}\le(v\ri)$ depends on the system and on the value of conserved densities, but it is manifestly independent of the details of the initial state. 

Intuitively, one can think of the slab of Minkowski spacetime of width $t$, in which the membrane is stretching, as a tensor network representation of the evolving state. In this analogy, the minimal membrane is a coarse grained minimal cut through the tensor network. For any tensor network, the number of legs cut by a minimal cut provides an upper bound on the entanglement entropy. In the coarse grained description, the membrane tension ${\cal E}\le(v\ri)$ is a proxy for the density of legs cut, and the minimal membrane computes the area instead of just providing an upper bound.
By reformulating the holographic entropy computation precisely in terms of the membrane theory, we hope to learn about the relation between tensor network descriptions, entanglement, and the dual gravitational dynamics~\cite{Swingle:2009bg,VanRaamsdonk:2010pw,Lashkari:2013koa,Maldacena:2013xja}. 

\section{Membrane picture from holography}

In recent work~\cite{Hartman:2013qma,Liu:2013iza,Liu:2013qca,Mezei:2016zxg} it was understood that in the small $\beta/R$ expansion, the leading order contribution to the entropy comes from the part of the extremal surface behind the horizon of the black brane created in a quench  or representing the dual geometry of to the thermofield double state that can also be used to compute the operator entanglement of the time evolution operator. Furthermore, it turns out that only the final state black brane geometry matters~\cite{Mezei:2016zxg}, the quench is accounted for only as a boundary condition on the extremal surface. For these reasons, we study the scaling limit of extremal surfaces in static black branes below.   

 The static geometry of the most general translation and rotation invariant asymptotically AdS black brane in infalling coordinates can be written as
 \es{MetricBH}{
ds^2&= {1\ov z^2}\le[-a(z)dt^2-{2\ov b(z)}\,dtdz+ d\vec{x}^2\ri]\,,
}
where the boundary is at radial position $z=0$, $\vec{x}$ is the field theory spatial coordinate, and $t$ is the infalling time. Without loss of generality we set the horizon location $z_h=1$, hence $a(1)=0$, and to get an asymptotically AdS spacetime we impose $a(0)=b(0)=1$. In terms of these data the two speeds associated with the spreading of quantum information are $v_E=\sqrt{-{a(z)\ov z^{2(d-1)}}}\big\vert_\text{max}$\footnote{max stands for the first local maximum behind the horizon, which we denote by $z_*$.} determining the early time linear growth of entanglement~\cite{Hartman:2013qma,Liu:2013iza,Liu:2013qca}, and $v_B=\sqrt{-{a'(1)\ov 2(d-1)}}$ related to out-of-time order correlators~\cite{Shenker:2013pqa,Roberts:2014isa}.

\begin{center}
\begin{figure}[!h]
\includegraphics[scale=0.35]{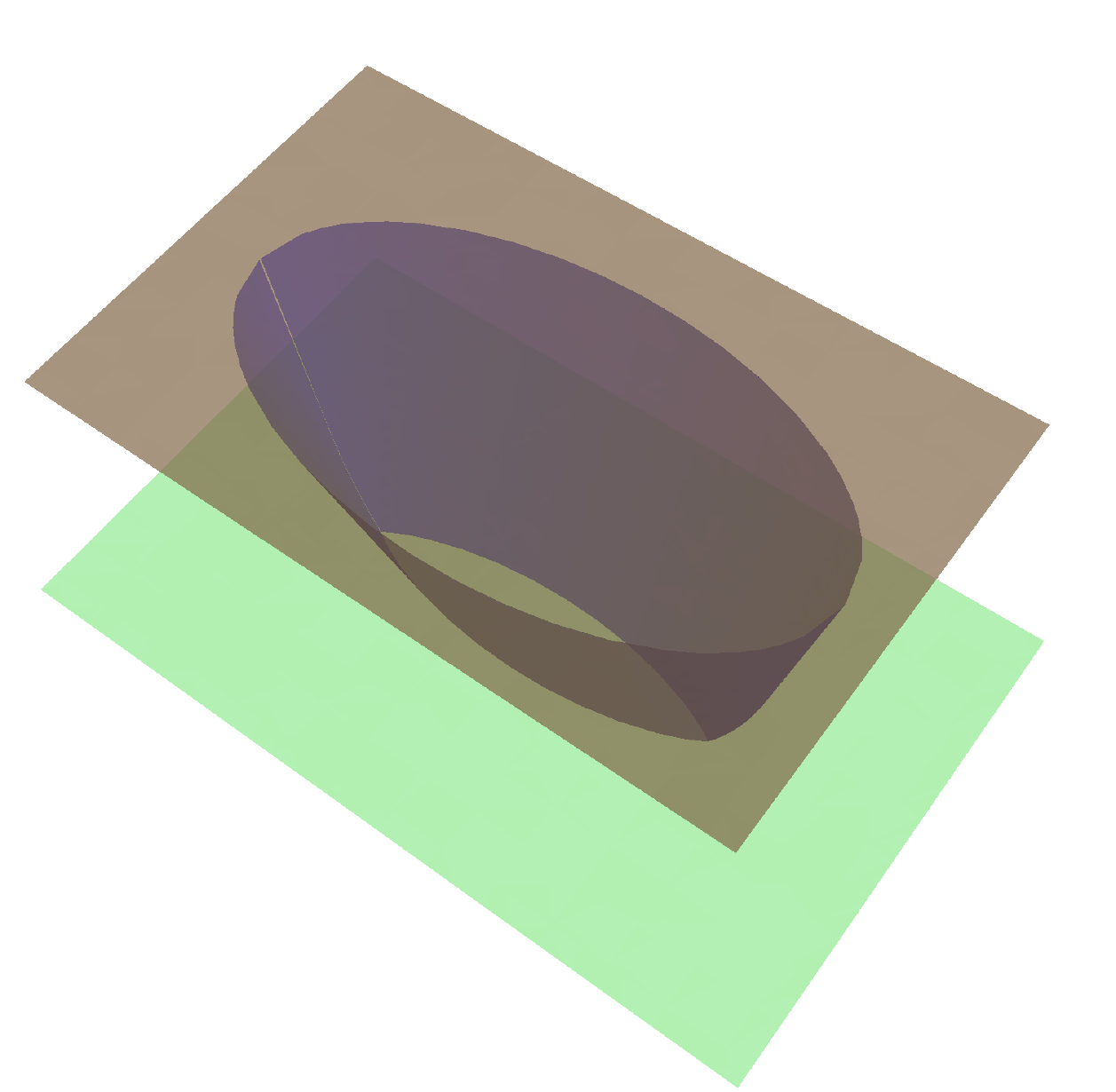}\hspace{0.75cm}
\includegraphics[scale=0.35]{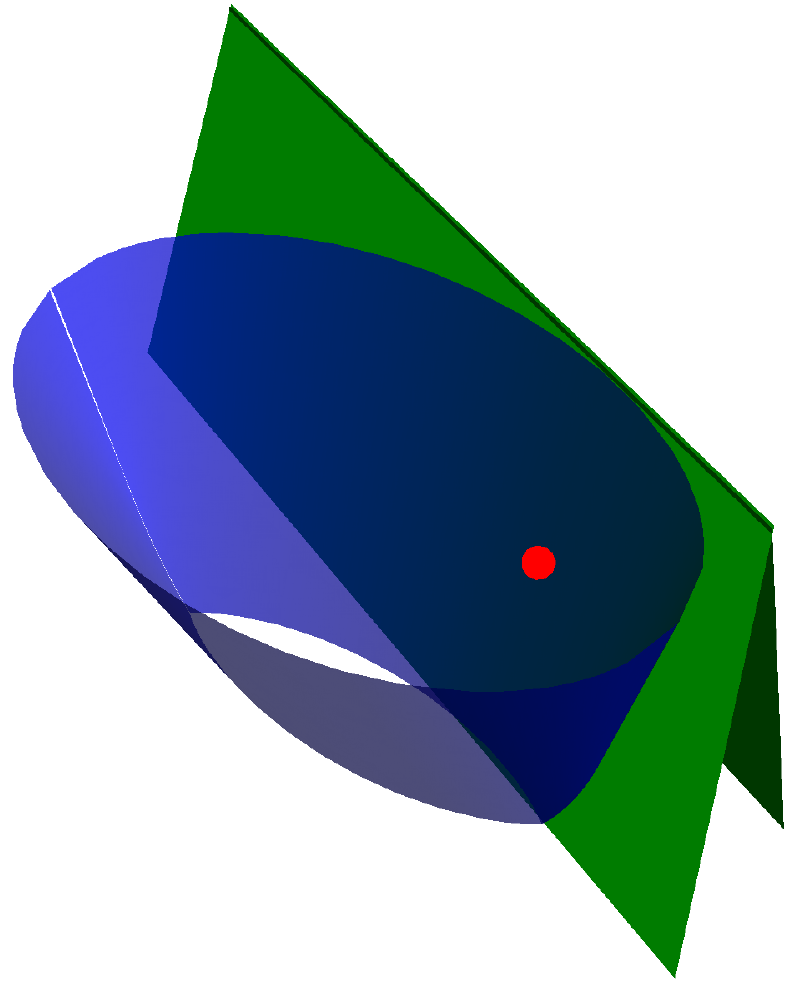} 
\caption{ {\it Left:}  A minimal membrane (blue) stretching between the two faces of a slab of Minkowski spacetime of width $t$. On the upper boundary it is an ellipse. {\it Right:} At the red point we drew a tangent plane (green), its angle with the vertical plane is $\arctan (v)$. For this membrane $v=v_B$ at every point, and we will refer to such membranes as ``light sheets" of slope $v_B$.
\label{fig:ellipse}}
\end{figure}
\end{center}

We want to understand extremal surfaces in the geometry \eqref{MetricBH} in a particular scaling limit corresponding to the regime \eqref{ScalingLimit} in field theory. It will be convenient to switch to polar coordinates $d\vec{x}^2=dr^2+r^2 d\Om^2$. We consider $t$ and the angular coordinates $\Om=\{\theta_1,\dots,\theta_{d-2}\}$ as independent variables, and the codimension-2 extremal surface will be given by two functions $r(t,\Om)$ and $z(t,\Om)$.\footnote{For complicated enough entangling surfaces these functions may not be single valued. In such cases we can work in coordinate patches and match these patches together to get the full surface.} It was understood in~\cite{Mezei:2016zxg} that to get the leading order entropy we have to implement the rescaling
\es{Scaling}{
t&\equiv R\,\tau\,, \quad\hspace{-0.1cm} r(t,\Om)  \equiv R \, \rho(\tau,\Om) \,,\quad \hspace{-0.1cm}
 z(t,\Om)  \equiv \zeta(\tau,\Om)\,.
}
In the limit $\beta/R\to 0$ the action simplifies, and takes the form in the scaled variables:
\es{AreaFunctScaled}{
S&=s_\text{th} R^{d-1} \int d\tau d\Om\ {\rho^{d-2}\ov \zeta^{d-1}}\sqrt{Q}\,,\\
Q&\equiv (\p_\tau \rho)^2- a(\zeta)\le(1+{(\p_\Om \rho)^2\ov \rho^2}\ri)\,,
} 
where $(\p_\Om \rho)^2$ is a shorthand for $g^{ij}_{S^{d-2}}(\p_{\theta_i}\rho)(\p_{\theta_j}\rho)$ with $g^{ij}_{S^{d-2}}$ the metric on a unit $S^{d-2}$.\footnote{The entropy in quantum field theory is divergent, and here we are implicitly working with the entropy from which we  subtracted the vacuum contribution to make it finite. Because the entropy obeys area law in the vacuum, for a finite cutoff the subtracted term can be regarded to be subleading in $\beta/R$. }
Note that the large $R$ limit brings major simplification: no derivatives of $\zeta$ appear in the action. Hence the equation of motion for $\zeta$ is algebraic:
\es{ZEOM}{
{(\p_\tau \rho)^2\ov 1+{(\p_\Om \rho)^2\ov \rho^2}}&=a(\zeta)-{\zeta a'(\zeta)\ov 2(d-1)} \,.
}
We make the definitions $\text{LHS}\equiv v^2$ and $\text{RHS}\equiv c(\zeta)$, with these new notations \eqref{ZEOM} takes the form $v^2=c(\zeta)$. Assuming the Null Energy Condition,
$c(\zeta)$ is a monotonically decreasing function until its first zero $\zeta_*$ as shown in Appendix~\ref{app:technical}. Because the LHS  is positive semidefinite, the relevant regime of $\zeta$ is  $0\leq \zeta \leq \zeta_*$, where 
 \eqref{ZEOM} can be symbolically inverted. Plugging back into the action the solution of the $\zeta$ equation of motion, $\zeta=c^{-1}(v^2)$, we can rewrite the extremization problem \eqref{AreaFunctScaled} as a minimal membrane problem of the form proposed in~\cite{Jonay:2018yei} with action:
\es{AreaFunctScaled2}{
S&=s_\text{th} R^{d-1} \int d\tau d\Om\ \rho^{d-2}\sqrt{1+{(\p_\Om \rho)^2\ov \rho^2}}\,{\cal E}\le(v\ri)\,, \\
&=s_\text{th} R^{d-1} \int d\text{area}\ { {\cal E}\le(v\ri)\ov \sqrt{1-v^2}}\,,
 }
 where the ``speed" $v={\p_\tau \rho/ \sqrt{1+{(\p_\Om \rho)^2\ov \rho^2}}}$ was defined below \eqref{ZEOM}, see also Fig.~\ref{fig:ellipse} for the geometric meaning of $v$ for the membrane.\footnote{Note that the $\sqrt{1-v^2}$ factor in the second line accounts for the fact that the area element in Minkowski spacetime in coordinates $(\tau,\,\Om)$ is $\rho^{d-2}\sqrt{1+{(\p_\Om \rho)^2\ov \rho^2}-(\p_\tau \rho)^2}=\rho^{d-2}\sqrt{1+{(\p_\Om \rho)^2\ov \rho^2}}\, \sqrt{1-v^2}$.} The membrane tension is given by
 \es{MembraneTension}{
 {\cal E}\le(v\ri)&\equiv{\sqrt{- a'(\zeta)\ov 2(d-1)\zeta^{2d-3}}}\Bigg\vert_{\zeta=c^{-1}(v^2)}\,.
} 
 Note that ${\cal E}\le(v\ri)$ is independent of the metric function $b(\zeta)$ in \eqref{MetricBH}, hence different spacetimes can give rise to the same entropy dynamics. We emphasize that by solving the algebraic equation of motion \eqref{ZEOM}, we reduced the holographic  extremization problem of a spacelike codimesion-2 surface in curved space, to a minimization problem of a timelike codimension-1 surface in Minkowski spacetime.\footnote{That we can talk about minimal (instead of extremal) membranes in Minkowski spacetime is a consequence of the form of  the membrane tension function to be discussed below: for constant membrane tension ${\cal E}_0$, for which ${\cal E}\le(v\ri)={\cal E}_0\sqrt{1-v^2}$ there does not exist a timelike minimal area membrane because of the zero area of lightlike membrane segments.}  The minimization problem does not make reference to the holographic extra dimension, the black brane spacetime has been repackaged into the membrane tension function ${\cal E}\le(v\ri)$. 
Recasting the entropy as a minimization problem, through the familiar holographic argument~\cite{Headrick:2007km}  immediately implies that strong subadditivity is satisfied.

${\cal E}\le(v\ri)$ is an even function of $v$. In the speed regime $0\leq v\leq v_B$, which as we discuss below is the dynamically relevant one,
 \es{Eprops}{
 {\cal E}\le(0\ri)&=v_E\,, \quad {\cal E}'\le(0\ri)=0\,, \quad {\cal E}\le(v_B\ri)=v_B\,, \quad {\cal E}'\le(v_B\ri)=1\,,\\
  {\cal E}'\le(v\ri)&\geq0\,,  \hspace{.5cm} {\cal E}''\le(v\ri)\geq0\,.
 }
${\cal E}\le(v\ri)$ is also defined for $ v_B<\abs{v}\leq 1$, but curiously it is not necessarily convex in that regime. For $v>v_B$ it obeys the following properties
  \es{Eprops2}{
 {\cal E}\le(1\ri)=\infty\,,\quad  {\cal E}'\le(v\ri)&>0\,, \quad {\cal E}'\le(v\ri) v-{\cal E}\le(v\ri)>0\,.
 }
 Note that it follows from \eqref{Eprops} that ${\cal E}'\le(v\ri) v-{\cal E}\le(v\ri)\leq0$ for $\abs{v}\leq v_B$, for example ${\cal E}\le(v\ri)$ functions that manifestly obey these properties see Fig.~\ref{fig:sketch}. For a proof of \eqref{Eprops} and  \eqref{Eprops2}, see Appendix~\ref{app:technical}. The convexity and the special behavior for $v=0,\,v_B$  were argued for on physical grounds in~\cite{Jonay:2018yei}; in this paper we provide a holographic proof of them and point out possible convexity violations for $\abs{v}>v_B$.

In a holographic conformal field theory for an equilibrium state with zero chemical potential,  
\es{EvSimp}{
{\cal E}\le(v\ri)&={v_E\ov (1-v^2)^{(d-2)/(2d)}}\,,\quad v_E={\le(d-2\ov d\ri)^{(d-2)/(2d)}\ov\le(2(d-1)\ov d\ri)^{(d-1)/d}}\,.
}
$ {\cal E}\le(v\ri)$ is plotted in Fig.~\ref{fig:sketch} for different values of the chemical potential. Note that the $d=2$ limit of \eqref{EvSimp} is  ${\cal E}\le(v\ri)=1$.
\begin{center}
\begin{figure}[!h]
\includegraphics[scale=0.6]{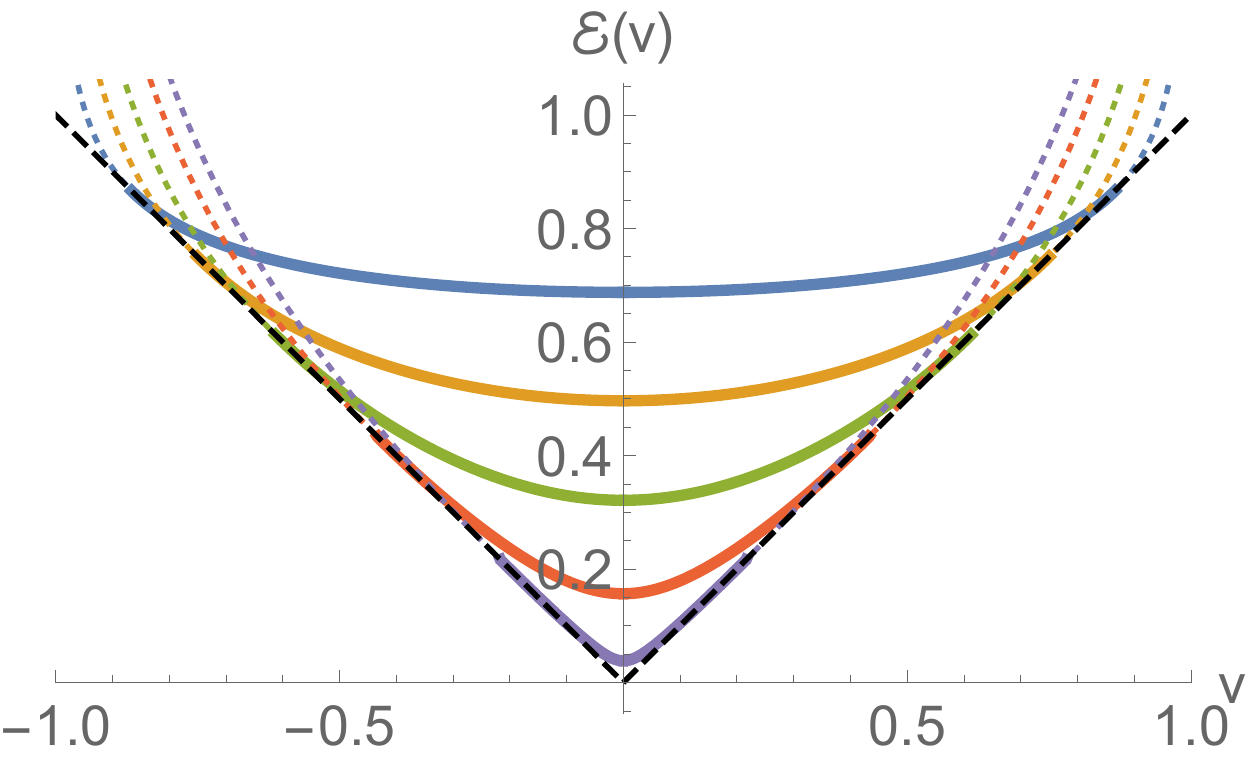} 
\caption{${\cal E}\le(v\ri)$ for different values of the chemical potential in $d=3$. The black brane metrics are given by \eqref{MetricBH}, with  $a(z)=1-(1+3q)z^3+3q z^{4}$ and $b(z)=1$, where the dimensionless boundary theory chemical potential is ${\beta\,\mu}\propto {q / (1-q)}$. We chose  $q=0,1/4,1/2,3/4,15/16$ for this plot, larger $q$ corresponds to smaller $v_E={\cal E}\le(0\ri)$. The black dashed lines are at $45^\circ$, at $v=v_B$, ${\cal E}\le(v\ri)$ by \eqref{Eprops} is equal to and tangent to this line. For $v\leq v_B$ we use solid, for $v>v_B$ dotted lines, as the latter regime is not important for the dynamics. 
\label{fig:sketch}}
\end{figure}
\end{center}

The above discussion misses static extremal surfaces that in the scaling limit $\beta/R\to 0$ lie  on the horizon $\zeta=1$, because these surfaces remain at a constant value of $\tau$, and one has to use $\rho$ and $\Om$ as independent variables to describe them. They can be incorporated into the membrane theory as an extra point with $v=\infty$ and  ${{\cal E}\le(v\ri)\ov\sqrt{1-v^2}}\big\vert_{v=\infty}=1$,\footnote{For such membranes we have to use the second line of \eqref{AreaFunctScaled2} to calculate the entropy as they do not extend in time.}  costing the same amount per unit horizontally projected area as a section of slope $v_B$. Horizontal sections of the membrane can be connected with sections with finite $v$.

\section{Dynamics, boundary conditions, and applications}

We now show that minimal membranes obey $\abs{v}\leq v_B$, and hence ${\cal E}\le(v\ri)$ for $\abs{v}>v_B$ is unimportant for the dynamics, as was noted in~\cite{Jonay:2018yei}. Tracing back through the formulas,\footnote{Note from \eqref{ZEOM} that $v=v_B$ corresponds to $\zeta=1$, hence $v\leq v_B$ means $1\leq \zeta$.} this means that the scaling extremal surfaces stay behind the horizon forever.

\begin{center}
\begin{figure}[!h]
\includegraphics[scale=0.4]{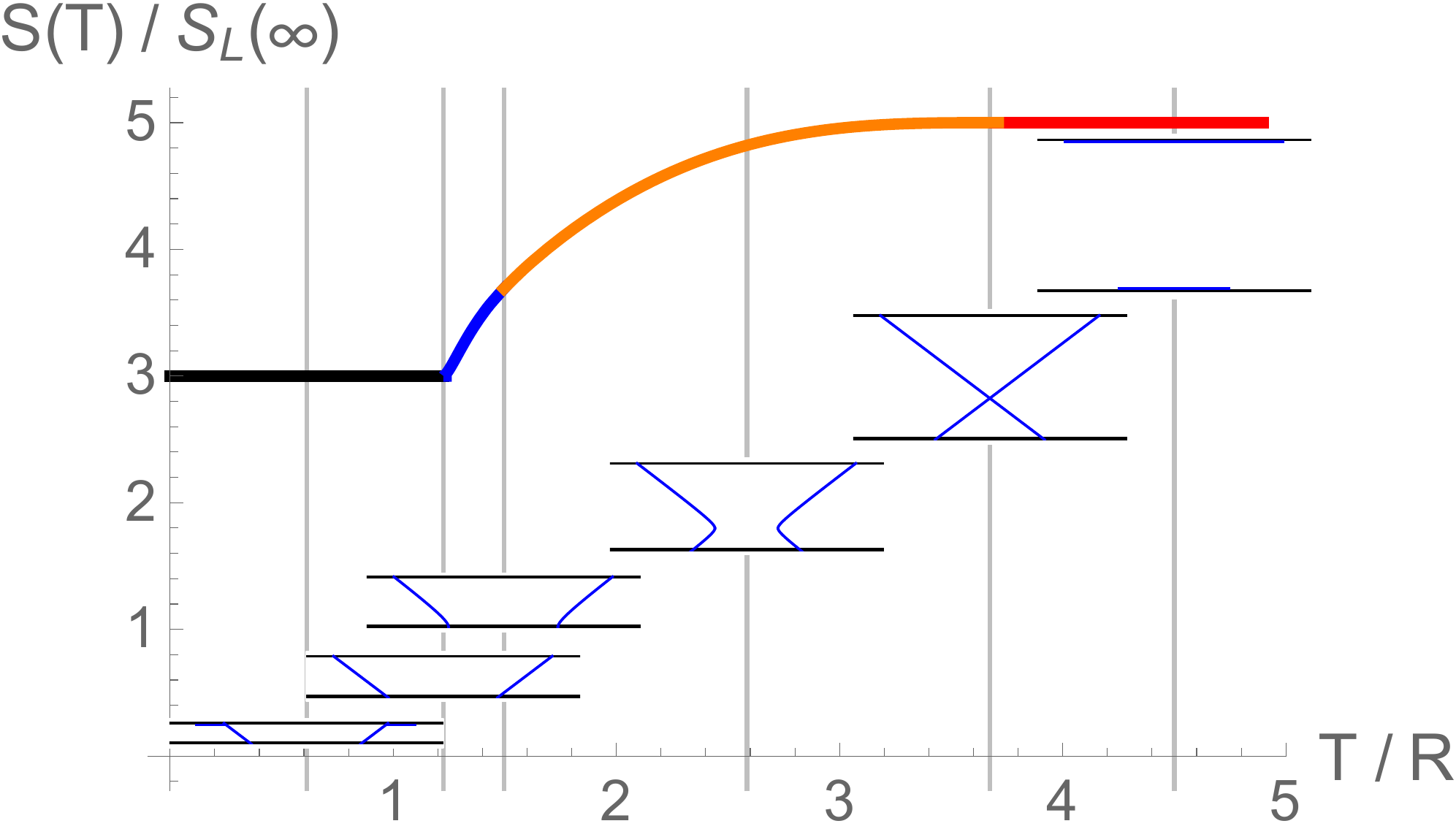} 
\caption{ Entropy growth in the thermofield double state in $d=4$ for two concentric spheres with $R_R/R_L=2$ as a function of $ T$. The entropy is normalized by the thermal entropy of the left sphere. Representative snapshots of membranes (blue) and the boundaries of the slab (black) are included at different times marked with gray gridlines. There are four regimes drawn by different colors matching the color coding used in~\cite{Mezei:2016zxg}. In the black regime the membrane consists of a horizontal and a ``light sheet" section that has slope $v_B$.  
In the blue section the membrane has increasing cross section as it interpolates between $L$ and $R$, while in the orange section it reaches a minimum in between. Finally, it transitions through a ``light sheet" of slope $v_B$ to a disconnected membrane in the red region. 
\label{fig:4dtwoside}}
\end{figure}
\end{center}

\begin{center}
\begin{figure}[!h]
\includegraphics[scale=0.4]{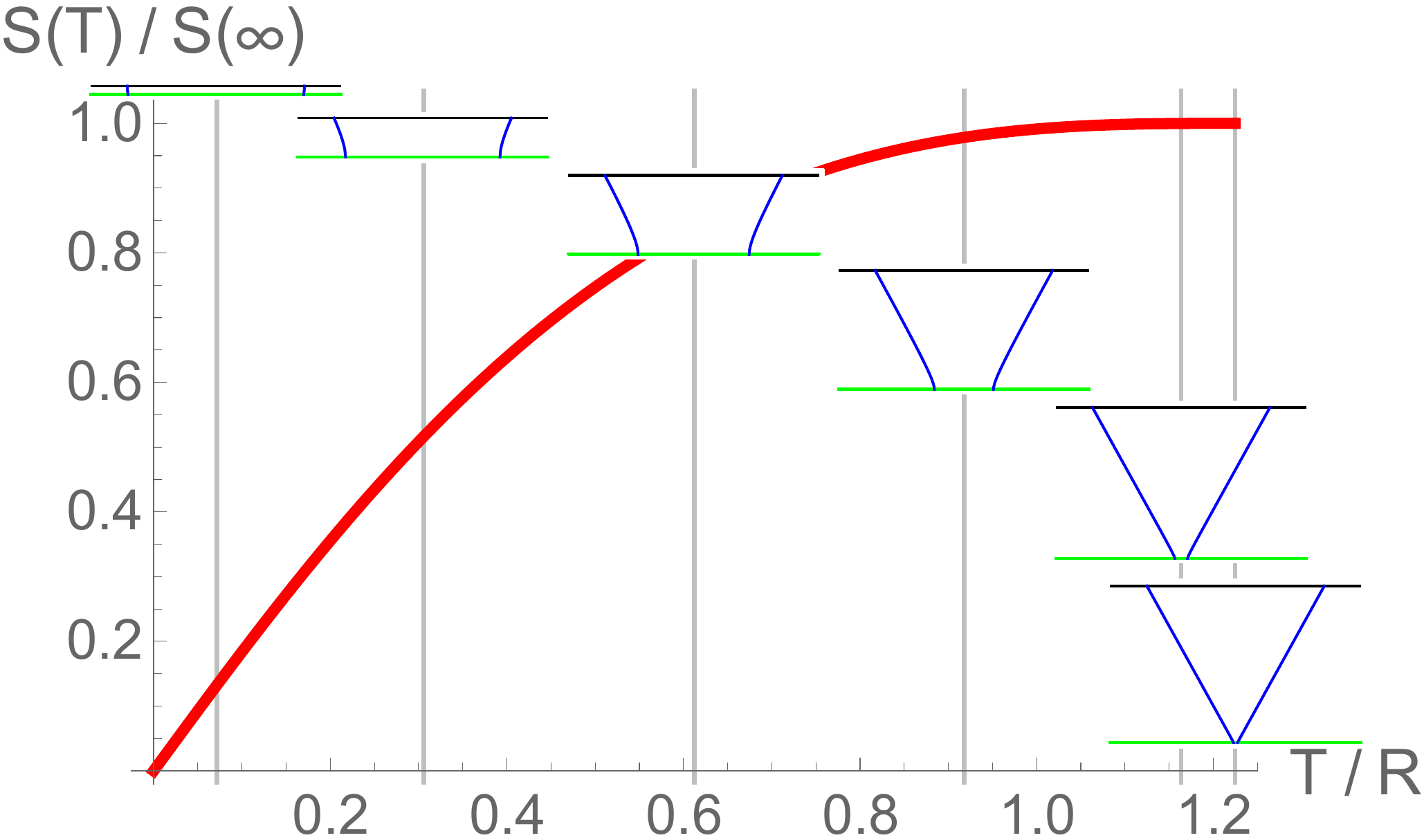} 
\caption{ Entropy growth in a charge neutral quench in $d=4$. Representative membranes shapes are included at  different times marked with gray gridlines. The membranes end perpendicular to the lower green end of the time slab, which represents the initial state. Note that at saturation, $T=R/v_B$ the membrane forms a cone.
\label{fig:4dquench}}
\end{figure}
\end{center}

Let us consider a minimal surface with spherical symmetry. Generic shapes are treated in Appendix~\ref{app:shape}. Because the Lagrangian \eqref{AreaFunctScaled2} is independent of $\tau$, we have a conserved energy,
\es{ConsEnergy}{
{H\ov s_\text{th} \,\text{area}(\p A)}=\rho^{d-2}\le[ {\cal E}'\le(v\ri) v-{\cal E}\le(v\ri)\ri]\,, \quad v(\tau)=\rho'(\tau)\,.
}
If at $\tau_0$ $v(\tau_0)\leq v_B$, then using the properties \eqref{Eprops} and \eqref{Eprops2}, conservation of energy implies that $v(\tau)\leq v_B$ for all times. Just as for any conservative one-dimensional mechanics problem, \eqref{ConsEnergy} allows for the solution of the minimal surfaces, and the determination of the entropy, see Figs.~\ref{fig:4dtwoside},~\ref{fig:4dquench} for examples. 

The scaling surfaces do not reach the boundary of AdS space. The complete extremal surfaces in the scaling limit consist of a part that skims the scaling surface (given by the membrane theory) and another that shoots out to the boundary in a way that only the radial coordinate, $\zeta$ changes, while $\rho$ and $\tau$ stay constant~\cite{Mezei:2016zxg}: this latter portion can be understood as an unstable mode around the scaling surface and does not contribute to the leading order entropy. Thus, on the scaling surface we simply have to impose the boundary condition
\es{ScalingBC}{
 \rho\le(\tau_f,\Om\ri) &=\rho_\text{bdy}(\Om)\,,
}
where $T=R\, \tau_f$ is the time at which we want to compute the entanglement entropy across the entangling surface given in polar coordinates by $r_\text{bdy}(\Om)=R\,\rho_\text{bdy}(\Om)$. The knowledge of the boundary condition \eqref{ScalingBC} allows us to compute the two-sided entanglement entropy in the thermofield double state dual to the AdS eternal black brane~\cite{Maldacena:2001kr,Hartman:2013qma}, where the entangling region consists of  $(0, r_\text{bdy,L}(\Om))\cup  (T, r_\text{bdy,R}(\Om))$: we have to compute the minimal membrane area stretching in a slab of width $ \tau={T/ R}$ between $ \rho_\text{bdy,L}(\Om)$ on the lower and $ \rho_\text{bdy,R}(\Om)$ on the upper boundary. This computation is equivalent to the computation of the operator entanglement entropy of $\sO=e^{-iH ( T -i\beta/2)}$~\cite{Hosur:2015ylk}. For two half-spaces offset by $ X$ we get~\cite{Jonay:2018yei}:
\es{HalfSpaces}{
S(T, X)=s_\text{th}\,\text{area}(\p A)\,  \begin{cases}
T\, {\cal E}\le({ X\ov T}\ri)  &( X\leq v_B T)\,,\\
X\quad &( X> v_B T)\,,
\end{cases}
} 
where the first line gives the area of flat membranes with $v= { X/  T}$, while the second line is the area of a membrane composed of a part with slope $v_B$ and horizontal segments.\footnote{There are many such membranes all giving equal area. At subleading orders in ${ \beta/ R}$ this degeneracy is lifted, and a membrane with reflection symmetry is the minimal membrane.}
For spheres the result follows from \eqref{ConsEnergy}, see  Fig.~\ref{fig:4dtwoside} for an example first solved in~\cite{Mezei:2016zxg}. 

In a quench, a short range entangled initial state is prepared and time evolved; by holography a quench is dual to black brane formation from collapse. Fast local thermalization in holography implies that the resulting spacetime only differs from the equilibrium black brane \eqref{MetricBH} for times $t=O(1)$ (at $z=O(1)$), and taking the scaling limit  \eqref{Scaling} reduces these geometries to the equilibrium black brane with a boundary at $\tau=0$. Extremal surfaces computing the entanglement entropy obey Neumann boundary conditions,
\es{ScalingBC2}{
\p_\tau \rho\le(\tau,\Om\ri) \vert_{\tau=0}&=0
}
on this boundary.  In the end of the world brane quench \cite{Hartman:2013qma} (dual to the boundary state quench model \cite{Calabrese:2005in}) this can be obtained by scaling the arbitrary boundary conditions one could impose on the brane~\cite{Mezei:2016zxg}.\footnote{The result remains true for more general branes considered in \cite{Kourkoulou:2017zaj,Almheiri:2018ijj}.} If the quench protocol instead involves dumping an incoherent energy density into the system in its vacuum state, the dual geometry is a collapsing shell of matter. For $\tau<0$ the geometry is that of pure AdS, and an explicit matching procedure can be executed for the extremal surface~\cite{Liu:2013iza,Liu:2013qca}, which after scaling  again leads to \eqref{ScalingBC2}~\cite{Mezei:2016zxg}. As the $\tau<0$ geometry represents the short range entangled vacuum, the contribution from this part of the extremal surface can be dropped at leading order in ${\beta/ R}$, and we simply cut the membrane at the $\tau=0$ wall. We conclude that there is universality in entropy growth in quenches: no matter what initial short range entangled state (or quench protocol) we choose, the chaotic evolution will erase the memory of it, and give rise to the same time evolution of entanglement entropy. The membrane tension function  ${\cal E}\le(v\ri)$ is manifestly independent of the details of the initial state and only depends on the final equilibrium state fixed by the value of the conserved charges.  For a strip geometry the entropy after a quench grows linearly with slope $s_\text{th}\, v_E\, \text{area}(\p A)$. For a spherical geometry we show a representative result in Fig.~\ref{fig:4dquench} obtained  using  \eqref{ConsEnergy}.

Another application of the membrane reformulation of the holographic entropy computation is the holographic proof of an inequality proven for a quench for an arbitrary chaotic quantum system in~\cite{Mezei:2016wfz}:
\es{tsunami}{
S[A(t)]\leq S[A'(t')]+s_\text{th}\,\le(\vol(A)-\vol(A'(t'))\ri)\,,
}
where $A'(t')$ is obtained by intersecting a past pointing ``light sheet" of slope $v_B$ starting from $\p A(t)$ with the time slice at $t'$. Let us consider a candidate membrane that is a ``light sheet" of slope $v_B$ between the time slices $t'$ and $t$, and a true minimal membrane between $0$ and $t'$. This candidate membrane has area equal to the RHS of \eqref{tsunami}. Its area is by definition greater than equal to the true minimal area giving  $S[A(t)]$, hence \eqref{tsunami} follows. Another inequality was proposed in~\cite{Mezei:2016wfz},
\es{BoundIntro}{
{dS\ov dt}\leq s_\text{th}\, v_E\, \text{area}\le(r_\text{bdy}(\Om)\ri)\,. 
}
Using Hamilton-Jacobi theory, one can show that the entropy growth from the functional \eqref{AreaFunctScaled2} with boundary conditions \eqref{ScalingBC} and \eqref{ScalingBC2} obeys\footnote{This result was shown in~\cite{Mezei:2016zxg} for the more complicated Lagrangian \eqref{AreaFunctScaled}, the proof for \eqref{AreaFunctScaled2} is a straightforward adaptation of the derivation there.}
\es{HJresult}{
{dS\ov dt}= s_\text{th}\, v_E\, \text{area}\le(r_\text{im}(\Om)\ri)\,,
}
where $r_\text{im}(\Om)=r\le(\tau=0,\Om\ri)$ is the image of the membrane on the initial time slice.  Using that $\text{area}\le(r_\text{im}(\Om)\ri)< \text{area}\le(r_\text{bdy}(\Om)\ri)$ \eqref{BoundIntro} follows. 

There is another way of obtaining the combination of bounds \eqref{tsunami} and \eqref{BoundIntro}. Fixing $v_E$ and $v_B$, we get the maximum possible entropy growth in the membrane model, if we choose the membrane tension to be ${\cal E}_\text{max}\le(v\ri)=v_E+\le(1-{v_E\ov v_B}\ri)\abs{v}$ for $\abs{v}\leq v_B$.\footnote{While this membrane tension does not satisfy the derivative conditions in \eqref{Eprops}, we can get arbitrary close to ${\cal E}_\text{max}\le(v\ri)$ with an ${\cal E}\le(v\ri)$ that satisfies all the conditions.} The minimal membranes in the  ${\cal E}_\text{max}\le(v\ri)$ model are built from a ``light sheet" part of slope $v_B$ and a vertical tube, with the joining time determined by minimizing the total area.\footnote{This is easiest to see, if we smooth out ${\cal E}_\text{max}\le(v\ri)$ by a small amount.} From this, we can verify that the entropy growth from ${\cal E}_\text{max}\le(v\ri)$ saturates the combination of bounds \eqref{tsunami} and \eqref{BoundIntro}~\cite{Jonay:2018yei}, so they are obeyed in any membrane model with arbitrary ${\cal E}\le(v\ri)$.  See~\cite{Mezei:2016wfz} for the worked out example of the sphere. In analogy, there exists a minimal tension
\es{Emin}{
{\cal E}_\text{min}\le(v\ri)=\begin{cases}
v_E \quad & (\abs{v}\leq v_E)\,,\\
\abs{v} \quad & (v_E< \abs{v} \leq v_B)\,.
\end{cases}
}
The entropy growth from ${\cal E}_\text{min}\le(v\ri)$ results in minimal membranes that are ``light sheets" of slope $v_E$, and the entropy is given simply by  
\es{MinEnt}{
S_\text{min}[A(t)]=s_\text{th}\,\le(\vol(A)-\vol(\hat{A}(t))\ri)
}
where $\hat{A}(t)$ is obtained by intersecting a past pointing ``light sheet"  of slope $v_E$ starting from $\p A(t)$ with the initial time slice. $S_\text{min}[A(t)]\leq S[A(t)]$, hence in \eqref{MinEnt} we provided a simple
lower bound on the entropy for fixed $v_E$. One  consequence of these considerations is that the saturation time is bounded by the times it takes $\hat{A}(t)$ and $A'(t)$ to shrink to zero size respectively:
\es{SatTime}{
{R_\text{insc}\ov v_E}\leq t_S \leq {R_\text{insc}\ov v_B}\,,
}
where $R_\text{insc}$ is the radius of the largest inscribable ball inside the region $A$.\footnote{Strips always saturate the lower bound in \eqref{SatTime}; in fact they saturate \eqref{MinEnt} for all times. For many holographic ${\cal E}\le(v\ri)$ functions, spheres saturate the upper bound~\cite{Mezei:2016zxg,Mezei:2016wfz}.} 

\section{Conclusions and outlook}

The main result of this paper is the rewriting of the holographic results of~\cite{Mezei:2016zxg} about the dynamics of entanglement entropy in the scaling limit \eqref{ScalingLimit} into the membrane theory form proposed by~\cite{Jonay:2018yei}. The membrane tension function ${\cal E}\le(v\ri)$ is determined by the equilibrium state dual to a static black brane, and we proved its properties \eqref{Eprops} and \eqref{Eprops2}. We showed that the resulting entropy satisfies the bounds discussed in~\cite{Mezei:2016wfz}. Besides unifying high energy and condensed matter approaches to entanglement dynamics, this reformulation also has many advantages: it allows a proof of new bounds \eqref{MinEnt} and \eqref{SatTime}, should make the problem numerically more tractable for arbitrary shapes, and provides a more intuitive picture than the results of~\cite{Mezei:2016zxg} given in \eqref{AreaFunctScaled}.  

 An intriguing qualitative connection between the geometry of tensor networks and spatial slices  of dual bulk spacetimes has been pointed out in~\cite{Swingle:2009bg,Hartman:2013qma,Qi:2013caa,Roberts:2014isa}. The minimal membrane can be thought of as a coarse grained minimal cut through the tensor network representation of the evolving wave function~\cite{Jonay:2018yei}. Hence, by deriving the membrane theory from holography, we put the connection between holography and tensor networks on a firm footing. Together with the difficulties in interpreting the whole spacetime as a tensor network~\cite{Qi:2018shh}, our results suggest that only after reducing the time direction in spacetime does the tensor network description emerge. In our setup, extremal surfaces that compute the entanglement entropy for different shapes (but at the same time) do not lie on the same spacelike surface. Instead,  we obtained the tensor network like description by solving for the trajectory of the extremal surface in the timelike direction $z$ using~\eqref{ZEOM}, and obtained an effective geometry.   It would be interesting to see whether the geometry encoded in ${\cal E}\le(v\ri)$ determines quantities other than the entanglement entropy and whether the knowledge of some set of nonequilibrium observables determines  ${\cal E}\le(v\ri)$. An early indication that this may be the case is the appearance of  $v_B$, encoded in out-of-time order correlators~\cite{Shenker:2013pqa,Roberts:2014isa}, in the dynamics of entanglement.

In holography, it is useful to think about the bulk radial direction $z$ as the field theory renormalization scale~\cite{Susskind:1998dq,deBoer:1999tgo,Verlinde:1999xm,Heemskerk:2010hk,Faulkner:2010jy}: the horizon at $z=1$ should be identified with the $\beta$. The geometry behind the horizon should encode degrees of freedom at wavelengths longer than $\beta$. From the study of entanglement growth at the largest scales \eqref{ScalingLimit}, we conclude that the entire region $1\leq z \leq z_*$ is equally important, hence the foliation of spacetime according to field theory scale breaks down behind the horizon (where $z$ is now timelike).

Extending the membrane theory to inhomogeneous and slow quenches, studying the holographic results for operator  entanglement of local operators and joining quenches (for which the membrane proposal already exists~\cite{Jonay:2018yei}), incorporating finite coupling and ``finite $N$" corrections to the holographic result, and including ${\beta/ R}$ corrections are all exciting problems. Another promising future direction is to test the membrane theory in other examples, and to possibly derive it. A strong test of the membrane theory in $d=2$ is to study the operator entanglement of $\sO=e^{-iH ( T -i\beta/2)}$ for two intervals of length $L$ displaced by $X$, as in~\cite{Asplund:2015eha}. The membrane theory predicts maximal entanglement scrambling for all chaotic theories: for intervals separated by $X>L$ the entropy is predicted to be time independent. R\'enyi entropies (except for the entanglement entropy) in all two-dimensional conformal field theories (including chaotic ones) are known to violate this prediction~\cite{Asplund:2015eha}, and hence the membrane theory in its current form does not apply to R\'enyi entropies.\footnote{Random circuits instead do not show difference between how R\'enyi and entanglement entropies behave.} Following the philosophy of~\cite{Mezei:2016wfz}, the membrane theory could also turn out to provide only an upper bound on the entropy in generic chaotic systems, then random circuits and holographic theories could be special systems saturating the upper bounds.

\section*{Acknowledgments}

I thank C.~Jonay, D.~Huse, and A.~Nahum for sharing a draft of their paper~\cite{Jonay:2018yei} prior to publication and useful discussions, and  T.~Hartman, H.~Liu, D.~Roberts, and D.~Stanford  for helpful discussions at various stages of the project. I also thank C.~Jonay, A.~Nahum, O.~Parrikar, D.~Stanford,  W.~van der Schee, and J.~Virrueta for detailed comments on the manuscript.
I am supported by the Simons Center for Geometry and Physics.

\appendix
\section{Technical results about $c(\zeta)$ and ${\cal E}\le(v\ri)$}\label{app:technical}

{\it Preparation:} In the Supplemental Material we switch notation $\zeta\to z$.  
We assume that the black brane geometry \eqref{MetricBH} obeys the Null Energy Condition (NEC),which implies that
\es{NEC}{
{d\ov dz}\le(b(z) a'(z)\ov z^{d-1}\ri)&\equiv s(z)\geq0\,,\\
b'(z)&\equiv t(z)\geq0\,,
}
where we defined the positive functions $s(z)$ and $t(z)$. We follow the proof technique developed in~\cite{Mezei:2016zxg} to write the metric functions $a(z)$ and $b(z)$ in terms of these sources:
\es{abSol}{
b(z)=&1+\int_0^z dz' \ t(z')\,,\\
a(z)=&\le(1-{B(z)\ov B(1)}\ri)\le(1-\int_0^1 dz' \ B(z') s(z')\ri)\\
&+\int_1^z dz' \ \le(B(z)-B(z')\ri) s(z')\,,
}
where we defined $B(z)\equiv\int_0^z dz'\ {z'^{d-1}\ov b(z')}$. It is easy to check that \eqref{abSol} satisfy the differential equations \eqref{NEC} and the boundary conditions discussed below \eqref{MetricBH}.

{\it Properties of $c(z)$:} We denote the first zero of $c(z)$ by $z_*$.\footnote{Note that $z_*$ is where the function $\sqrt{-{a(z)\ov z^{2(d-1)}}}$ reaches its maximum value $v_E$. This is the location of the Hartman-Maldacena surface~\cite{Hartman:2013qma}.} We prove below that in the regime relevant for the membrane theory,  $0\leq z \leq z_*$,  $c(z)$ is invertible, because it is  monotonically decreasing from $c(0)=1$, through $c(1)=v_B^2$, to $c(z_*)=0$.

Let us write:
\es{cmono}{
&c'(z)=-{1\ov 2(d-1)}\le(z\,a''(z)-(2d-3)a'(z)\ri)\\
&=-{1\ov 2(d-1)} \le[{z^{d-2}\ov b(z)}s(z)+\le(d-2+{t(z)\ov z\, b(z)}\ri)(-a'(z))\ri]\,,
}   
where we used the definition of $c(z)$ given below \eqref{ZEOM} and the NEC inequalities \eqref{NEC} (written as equalities). $c'(z)\leq 0$ would follow, if $-a'(z)\geq 0$ were true, as then all  terms in products above would be positive. 

It is easy to prove that $a'(z)\leq 0$ for $1\leq z \leq z_*$.  
The positivity of $c(z)$ is equivalent to
\es{apBehind}{
a'(z)\leq {2(d-1)a(z)\ov z}\leq0\,, 
}
 where we used that $a(z)\leq0$ behind the horizon. Thus, we established $a'(z)\leq 0$ in this regime.

For $0\leq z < 1$ the proof has a different flavor. Using \eqref{abSol}, we can write
\es{aprime}{
a'(z)=&-B'(z)\le[{1\ov B(1)}\,\le(1-\int_0^1 dz' \ B(z') s(z')\ri)\ri.\\
&\le.+\int_z^1 dz' s(z')\ri]\,.
}
We now have to use the fact that there is no horizon in the range $0\leq z < 1$, hence $a'(1)<0$  has to hold. Plugging $z=1$ into \eqref{aprime} then implies   $\int_0^1 dz' \ B(z') s(z')<1$, which then establishes $a'(z)<0$ outside the horizon. We conclude that $c(z)$ is monotonically decreasing on the entire  $0\leq z \leq z_*$ interval of interest.

{\it Properties of ${\cal E}\le(v\ri)$:} We now turn our attention to showing \eqref{Eprops} and \eqref{Eprops2}.  To obtain information about the derivatives of ${\cal E}\le(v\ri)$ we compute:
\es{keyresult}{
{\cal E}'\le(v\ri)= {1\ov z^{d-1}}\,\sqrt{1-{2(d-1) a(z)\ov z \,a'(z)}}\Bigg\vert_{z=c^{-1}(v^2)}\,, \quad (v\geq0)\,.
}
The positivity of the expression under the square root follows from $a'(z)<0$ outside the horizon, and from \eqref{apBehind}  for $1\leq z \leq z_*$. ${\cal E}'\le(v\ri)$  is manifestly positive. The first line of \eqref{Eprops} follows straightforwardly using that $v=0,\,v_B$ corresponds to $z=z_*,\,1$ respectively. For $v>v_B$, corresponding to $0\leq z<1$ 
\es{keyresultB}{
{\cal E}'\le(v\ri)>1\,, \quad (v>v_B)\,,
}
as in \eqref{keyresult} it is given by a product of two numbers both greater than 1.
To investigate the convexity properties of  ${\cal E}\le(v\ri)$ we compute for $ v\geq0$: 
\es{keyresult2}{
{\cal E}''\le(v\ri)=&{\le[2(d-1)/(-a'(z))\ri]^{3/2}\ov z^{d+1/2}\, \, (-c'(z))}\\
&\times \le[(-a(z))(-c'(z))+2(-a'(z))c(z)\ri]\big\vert_{z=c^{-1}(v^2)}\,,
}
where the prefactor is positive. For $1\leq z \leq z_*$ both terms in the square bracket are positive, because they are products of positive terms. Hence we established ${\cal E}''\le(v\ri)$ for $ \abs{v}\leq v_B$. For $0\leq z<1$ the proof breaks down, and we can find counterexamples where ${\cal E}''\le(v\ri)<0$: let us take $t(z)=0$ and $s(z)={\ep\ov z_0^d} \delta(z-z_0)$, then 
\es{trouble}{
&\le[(-a(z))(-c'(z))+2(-a'(z))c(z)\ri]\\
&=-{(1-z_0^d)(d-\ep)\ov 2 d(d-1)} \ep\, \delta(z-z_0)+\text{const}\,,
}
which for small $\ep$ is negative, taking ${\cal E}''\le(v\ri)$ negative.\footnote{A potential worry in postulating and arbitrary $s(z)$ is that it can create a horizon at some $z<1$. Here we found an effect for small $\ep$, which only changes $a(z)$ by $O(\ep)$, hence we found a valid counterexample. We have also constructed examples with smooth $s(z)$, which lead to ${\cal E}''\le(v\ri)<0$.}

Finally, we demonstrate that for $v>v_B$,  ${\cal E}'\le(v\ri) v>{\cal E}\le(v\ri)$. Equivalently, we want to show that the following combination is positive definite for $0\leq z <1$:
\bwt
\es{squarediff}{
\le[{\cal E}'\le(v\ri) v\ri]^2-\le[{\cal E}\le(v\ri)\ri]^2&={2a(z)\ov z^{2d-1}\, B'(z)}\,\le[zB'(z)+(d-1)\,{B(1)\le(C+\int_z^1dz'\ B(z') s(z')\ri)-B(z)\le(C+\int_z^1dz'\ B(1)  s(z')\ri)\ov C+\int_z^1dz'\ B(1)  s(z')}\ri]\,,\\
C&\equiv 1-\int_0^1 dz' \ B(z') s(z')>0\,,
}
\ewt
where $C>0$ was shown below \eqref{aprime}, $a(z)>0$ in this regime of $z$, and $B(z)$ is a positive and monotonically increasing function. We then decrease the RHS by using $\int_z^1dz'\ B(z') s(z')>B(z)\int_z^1dz'\  s(z')$ and get
\bwt
\es{squarediff2}{
\le[{\cal E}'\le(v\ri) v\ri]^2-\le[{\cal E}\le(v\ri)\ri]^2&>{2a(z)\ov z^{2d-1}\, B'(z)}\,\le[zB'(z)+(d-1)\,{\le(B(1)-B(z)\ri)C \ov C+\int_z^1dz'\ B(1)  s(z')}\ri]\,,
}
\ewt
which is now manifestly positive. This concludes the proof of the relevant properties of ${\cal E}\le(v\ri)$.

\section{Minimal membranes have $\abs{v}\leq v_B$, or scaling surfaces do not cross the horizon }\label{app:shape}

Let us assume that there exists a membrane, which has a point whose slope $v$ crosses $v_B$ as a function of $\tau$. For simplicity, we will restrict to $d=3$. We series expand $\rho(\tau,\theta)$ as
\es{rhoSeries}{
&\rho(\tau,\theta)=\rho_0+v_B\, \tau+\le[a_{2,0}\tau^2+a_{1,1}\tau\theta+a_{0,2}\theta^2 \ri]\\
&+\le[a_{3,0}\tau^3+a_{2,1}\tau^2\theta+a_{1,2}\tau\theta^2+a_{0,3}\theta^3 \ri]+\dots\,,
}
where by choosing the origin of polar coordinates appropriately, we set $\theta=0$ at the candidate point and we got rid of the linear in $\theta$ term. If we compute $v$, we get
\es{vSeries}{
v(\tau,\theta)=&v_B+\le[2a_{2,0}\tau+a_{1,1}\theta \ri]+O\le((\tau,\theta)^2\ri)\,,
}
and requiring that at $\tau=0$ there is no point that has $v>v_B$ sets $a_{1,1}=0$. Having  $a_{2,0}>0$ gives a membrane that at $\theta=0$ has $v<v_B$ for $\tau<0$, and surpasses $v_B$ for $\tau>0$.  However, plugging into the equation of motion, we get that $a_{2,0}=a_{3,0}=a_{2,1}=0$. Thus \eqref{vSeries} reduces to
\pagebreak
\es{vSeries2}{
v(\tau,\theta)=&v_B+\le[a_{1,2}-{2v_B a_{0,2}^2\ov \rho_0^2} \ri]\theta^2+O\le((\tau,\theta)^3\ri)\,,
}
implying that instead of changing speed  the $\theta=0$ point stays at $v=v_B$. In fact the pure $\tau^3$ term also vanishes, and any other time dependent term appearing at higher orders cannot overcome the $\theta^2$ term, making it impossible for the speed to cross $v_B$ away from $\theta=0$ within the range of validity of the expansion.  One can perform this exercise to high orders without a change in conclusions.

What we found above is that the minimal surface locally become a ``light sheet" of slope $v_B$.  A ``light sheet" shot from any entangling surface is always minimal: the above perturbative discussion can be reinterpreted as a proof of this fact. One can also verify this by plugging an arbitrary ``light sheet" into the equation of motion. Minimal membranes and ``light sheets" in particular  in general become cuspy, due to the familiar phenomenon of swallowtail singularities in propagating wave fronts. See Figure~\ref{fig:ellipse} for an example of a ``light sheet" of slope $v_B$ that becomes cuspy at some finite depth in $\tau$. An important observation is that even for such a cusp on a ``light sheet" $v=v_B$, as can be easily verified from considering two planes of slope $v_B$ intersecting at an angle and using the definition of $v$ below \eqref{ZEOM}.

\bibliographystyle{ssg}
\bibliography{speeds}

\end{document}